\documentclass[12pt]{article}
\usepackage{amsmath}
\usepackage{graphicx}
\usepackage{enumerate}
\usepackage{natbib}
\usepackage{url} 

\usepackage{siunitx}
\usepackage{mathrsfs}
\usepackage{xcolor}

\providecommand{\noopsort}[1]{}
\newcommand{\blind}{1}

\newcommand{\de}{\mathrm{d}}

\begin{document}
\graphicspath{{pics/}{pics/}}

\def\spacingset#1{\renewcommand{\baselinestretch}%
{#1}\small\normalsize} \spacingset{1}


\if1\blind
{
  \title{\bf A multitype Fiksel interaction model for tumour immune microenvironments}
  \author{Jonatan A. Gonz\'alez \\
    Computer, Electrical and Mathematical Science and Engineering Division, \\
    King Abdullah University of Science and Technology (KAUST), \\
    Thuwal 23955-6900, Saudi Arabia\\
    and \\
    Paula Moraga \\
    Computer, Electrical and Mathematical Science and Engineering Division, \\
    King Abdullah University of Science and Technology (KAUST), \\
    Thuwal 23955-6900, Saudi Arabia\\}
  \maketitle
} \fi

\if0\blind
{
  \bigskip
  \bigskip
  \bigskip
  \begin{center}
    {\LARGE\bf A multitype Fiksel interaction model for tumour immune microenvironments}
\end{center}
  \medskip
} \fi

\bigskip
\begin{abstract}
The tumour microenvironment plays a fundamental role in understanding the development and progression of cancer. This paper proposes a novel spatial point process model that accounts for inhomogeneity and interaction to flexibly model a complex database of cells in the tumour immune microenvironments of a cohort of patients with non-small-cell lung cancer whose samples have been processed using digital pathology techniques. Specifically, an inhomogeneous multitype Gibbs point process model with an associated Fiksel-type interaction function is proposed. Estimation and inference procedures are conducted through maximum pseudolikelihood, considering replicated multitype point patterns.
\end{abstract}

\noindent%
\vfill

\newpage
\spacingset{1.9} 

\section{Introduction}
\label{sec:intro}
Tumours are complex ecosystems that consist of much more than a collection of cancer cells. For example, they contain epithelial cells, fibroblasts, blood and lymphatic vessels, and infiltrating hematopoietic cells, among others \citep{remark2015nonsmallcancer}. These structural elements affect the growth and clinical conditions of the tumour. The ecosystem surrounding a tumour within the body is usually known as the tumour microenvironment. It is a set of infiltrating and resident host cells, secreted factors, and extracellular matrix \citep{jhonson2021cancercellspecific}. Tumour cells stimulate essential molecular, cellular, and physical changes within host tissues to support tumour growth and progression. The composition of the tumour microenvironment varies between tumour types, but distinctive features include immune and stromal cells, among others. The tumour microenvironment characterises the tumour and its environment and plays a fundamental role in understanding the development and progression of cancer. One of the key components of the tumour microenvironment is the tumour immune microenvironment, which has a highly diverse composition, including various populations of T-cells, B-cells, dendritic cells, natural killer cells, myeloid-derived suppressor cells, neutrophils, or macrophages \citep{remark2015nonsmallcancer}.

Recent advances in imaging techniques allow scientists to study the spatial structure of the tumour microenvironment or tumour immune microenvironment at a level of detail down to a single cell \citep{digitalpathology2014}. This data (several antibody markers) has complex acquisition and processing. Processing them involves various efforts in the laboratory and applying various numerical and statistical methods to reduce non-biological variability, i.e., the variability due to the computational procedures to process the data \citep{harris2022multiplexslidetoslidevariation}. Some research teams have developed suitable methods and software to address these challenges. 

In this context, image normalisation is a technique that adjusts an image's input pixel- or image-level values to remove noise and improve image quality. Some statistical tools for normalisation improve the similarity across images by removing the unknown effect of technical variability. To normalise multiplex image data, \citep{harris2022multiplexslidetoslidevariation} implement and compare data transformations and normalisation algorithms in multiplexed imaging data providing a foundation for better data quality and evaluation criteria in multiplexed imaging. \citep{seal2022denVar} propose a density-based method for distinguishing the difference between the subjects concerning the distribution of a functional marker in the tumour microenvironment or tumour immune microenvironment. \citep{Steinhart2021SpatialOvarian} examine how spatial interactions among different immune cells in the ovarian cancer tumour microenvironment are associated with overall survival using scalar spatial summaries. 

Currently, spatial statistical techniques are preferred when analysing this type of data \citep{wilson2021oportunitiesmultiplexdata}. The locations of the immune cells in the tumour immune microenvironment can be assumed as a spatial point pattern in a predefined observation window, usually given by the limits in which the image of the biological sample was processed. The most straightforward point process is completely spatial random (CSR or stationary Poisson process), where the expected value of the number of immune cells is assumed to be constant throughout the region of interest and where the cells do not interact with each other \citep{diggle2013book}. This model, however, is unrealistic in practice since cells easily violate both assumptions \citep{diggle2006_amacrinebook}, accumulate in certain preferred regions of the tissue (inhomogeneity), repel or attract each other, or even attract each other on one scale and repel each other on another scale (interaction). Therefore, we can study the tumour microenvironment or tumour immune microenvironment from the spatial point processes point of view by employing some tools to deal with inhomogeneity and interaction. We consider the different cell sub-types within the tumour microenvironment or tumour immune microenvironment to provide helpful information on how cells behave and how their distribution is affected. We also consider various exogenous factors simultaneously. This could allow future medical or clinical decisions regarding the patient to be positively influenced by the knowledge acquired about this cellular dynamic.

Analysing densities and interactions between points in some spatial domain is a primary pursuit in spatial statistics \citep{baddeley2015spatialR}. Some real datasets have motivated these analyses; for example, biology \citep{harkness1983ants}, neuroscience \citep{diggle1986amacrine} and ecology \citep{stoyan2000applicationsforestry}. Commonly, the literature describes multivariate point patterns through second-order summary descriptors such as the $K$- or $J$- functions in their multitype versions \citep[see, e.g., \ ][]{lotwick:silverman:82, vanliesout1999mutivariatej}. There are other methods for multitype point patterns, such as the mark connection function more suitable for detecting mark correlation in an exploratory analysis \citep{baddeley2015spatialR}. Testing spatial independence between two components of a stationary bivariate spatial process is a well-known problem in the literature \citep{diggle2013book, tomas2020revisiting}. 

Gibbs point processes are a wide class that includes, for example, all Cox processes and all finite point processes having a density with respect to the Poisson process \citep[see, e.g., \ ][]{baddeley2015spatialR}. Gibbs processes are motivated by statistical physics and arise from the forces acting on and between particles in a fluid or gas. We start by assuming that the total potential energy $V(\cdot)$ corresponding to a given configuration of particles, that is, an instantaneous snapshot, can be disaggregated into different terms that represent the potential energies of the individual particles (which can come from external force fields), the interactions between particles taken in pairs, triples, etc. Often it is assumed that only the first and second-order terms need to be included. Then, a representation of the total potential energy for $n$ particles $X=\{\xi_i\}_{i=1}^n$ would be given by \cite{daley2003} $$V(X)=V(\xi_1,\ldots,\xi_n)=\sum_{j=1}^n \sum_{1\leq i_1< \cdots < i_j\leq n} V_j(\xi_{i_1},\ldots,\xi_{i_j}),$$
where $V(\cdot)$ is the interaction potential of order $j$. One of the principles of statistical mechanics establishes that, in equilibrium, the probability density of a point pattern, that is, a particular configuration of points, is inversely proportional to the exponential of the potential energy; that is, proportional to $e^{-V(X)/T}$, where $T$ is the temperature. Potential energy is the total work required to move the particles to form the point pattern $X$. Markov point processes, a virtual subclass of these Gibbs processes where the interaction range of particles is assumed to be finite, are flexible statistical models for spatial point patterns \citep[see, e.g., \ ][]{vanLieshout2000markov, diggle2006_amacrinebook}. 

In this paper, we propose a novel approach that leverages several spatial statistical techniques to model the distributions of cells in tumour immune microenvironments flexibly. We employ a non-small cell lung cancer (NSCLC) dataset collected by multiplex immunohistochemistry (mIHC) \citep{jhonson2021cancercellspecific}. The data provide tissue samples collected from 122 non-small cell lung cancer patients. These samples were processed to isolate the tumour immune microenvironments and obtain marked point patterns where each point represents an immune cell, which is marked as belonging to one of five immunity markers (see Section \ref{sec:data}). Similarly, the dataset includes some clinical factors such as age, whether the patient has undergone chemotherapy, the stage of the disease and survival time. Our main objective is to develop a multitype inhomogeneous point process model for the cells of the tumour immune microenvironment that includes the acquired contextual knowledge, i.e., including the marks of immunity, the clinical factors of the patients and the possible interaction between cells of the same and different types.

We formulate an accurate statistical model and validate it to answer the scientific question behind our research objective, taking advantage of all the data components. To do this, we start from a general principle of points interacting in Gibbs' fashion and their probability distribution. We incorporate several factors, such as the trend, whose baseline is estimated using non-parametric techniques. We add a multitype pairwise interaction component inspired by cell dynamics. We use several methods for estimation and inference: the descriptive input of second-order statistics such as Ripley's $K$-function, the profile pseudolikelihood, and the maximum pseudo-likelihood maximisation. In addition, since the data were collected from several patients, we take advantage of methods related to replicated point patterns to feed the model and gain more robust estimates of the model parameters.

The remainder of this article is organised as follows. We describe the tumour immune microenvironment dataset in Section \ref{sec:data}. Section \ref{sec:gibbsmodels} contains the fundamental notions about Gibbs's processes. In Section \ref{sec:fikselinteraction}, we introduce the Fiksel interaction function in its multitype version and describe the methods we utilise to make statistical inference. In Section \ref{sec:application}, we detail the analysis of the tumour immune microenvironment dataset step by step. We estimate the model's components by combining several techniques starting from the corresponding geometric considerations and also compare the model's performance with several other alternative models. We define a type of root mean square error (RMSE) based on residuals to facilitate comparisons. We end with some comments, directions for future research, and final considerations in Section \ref{sec:discussion}.

\section{Immune cells data}\label{sec:data}
The cellular composition of the tumour immune microenvironment can be studied through multiple well-known techniques in digital pathology \citep[see, e.g.,\ ][and references therein]{wilson2021oportunitiesmultiplexdata}. In this work, the data come from multiplex immunohistochemistry (mIHC), which allows the evaluation of multiple markers in a single experiment and may detect the spatial location of multiple cell types. 

\cite{jhonson2021cancercellspecific} used multispectral quantitative imaging on the lung adenocarcinoma tumour microenvironment in 153 patients with resected tumours. The data consist of a single slide per patient, where they evaluated the tumour microenvironment with markers for CD3, CD8, CD14, CD19, major histocompatibility complex II (MHCII), cytokeratin, and 40,6-diamidino-2-phenylindole (DAPI). Then they performed image analysis, including tissue segmentation, phenotyping, and attached spatial coordinates. The data are available at \cite{VectraPolarisData}.

\begin{figure}[htb]
	\centering
	\includegraphics[width=0.5\linewidth]{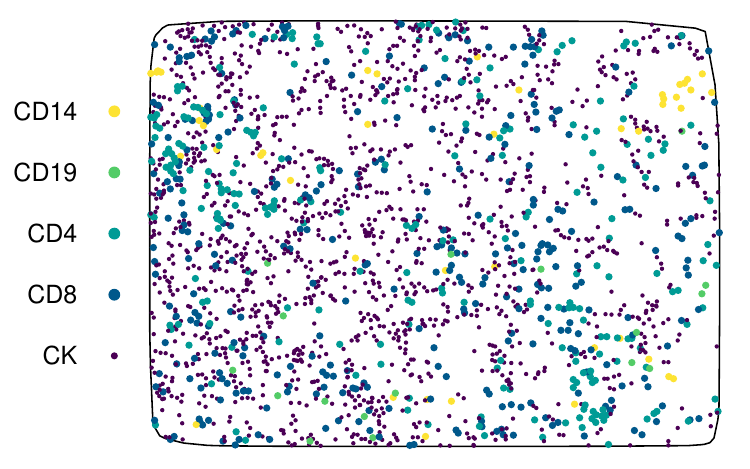}
	\caption{\label{fig:sample1} Representative phenotype map of the tumour immune microenvironment of a random patient.}
\end{figure}

Specifically, the data associated with each patient comes in a point pattern format representing the phenotype map of CK$^+$ cancer cells, CD4$^+$ (CD3$^+$CD8$^-$) T-cells, CD8$^+$ T-cells, CD14$^+$ cells and CD19$^+$ B-cells. A random (patient 45) processed tissue sample is displayed in Figure \ref{fig:sample1}. 
The data for this study came from the Mayo Clinic Lung Cancer Repository, where they ensured compliance with applicable ethical and data protection protocols. The selected patients (a total of 122) underwent curative surgical resection of lung adenocarcinoma between 2004 and 2007. These patients had not received targeted anticancer therapy and had available residual tumour specimens. 

In addition, extra information related to each patient was extracted, which will be considered design covariates (non-spatial). The covariates are, gender (56\% women), age at the time of surgery (a mean of 68 years), stage of cancer (42\% IA, 23\% IB, 10\% IIA, 10\% IIB, 12\% IIIA, 1\% IIIB, 3\% IV), cancer cell MHCII status (67\% high ($\geq 0.5\%$)), survival days (a mean of 2389 days, i.e., approx. six years and a half), death (44\% dead), recurrence or dead event (38\% of no recurrence), adjuvant therapy (86\% of no therapy). To explain the state of cancer, we follow \cite{lungcancerinstitute2023}. Stage I is divided into stages IA and IB. In stage IA, the tumour is only in the lung and up to \SI{3}{\cm}. At this stage, cancer has not spread to the lymph nodes. In stage IB, the tumour size lies between \SI{3}{} and \SI{4}{\cm}, and cancer has not spread to the lymph nodes. Stage II is also divided into two categories, IIA where the tumour lies between \SI{4}{} and \SI{5}{\cm} and cancer has not spread to the lymph nodes, and IIB, where the tumour lies between \SI{4}{} and \SI{5}{\cm} and cancer has spread to lymph nodes on the same side of the chest as the primary tumour; the lymph nodes with cancer are in the lung or near the bronchus. Stage III is divided into three categories, IIIA, where the tumour is up to \SI{5}{\cm} and cancer has spread to lymph nodes on the same side of the chest as the primary tumour; the lymph nodes with cancer are around the trachea or aorta, or where the trachea divides into the bronchi. In IIIB, the tumour is up to \SI{5}{\cm}, and cancer has spread to lymph nodes above the collarbone on the same side of the chest as the primary tumour or to any lymph nodes on the opposite side of the chest as the primary tumour. In IIIC, the tumour may be any size, and cancer has spread to lymph nodes above the collarbone on the same side of the chest as the primary tumour or to any lymph nodes on the opposite side of the chest as the primary tumour. Finally, stage IV, where the tumour may be any size and cancer may have spread to the lymph nodes. 

\begin{figure}[htb]
	\centering
	\includegraphics[width=0.24\linewidth]{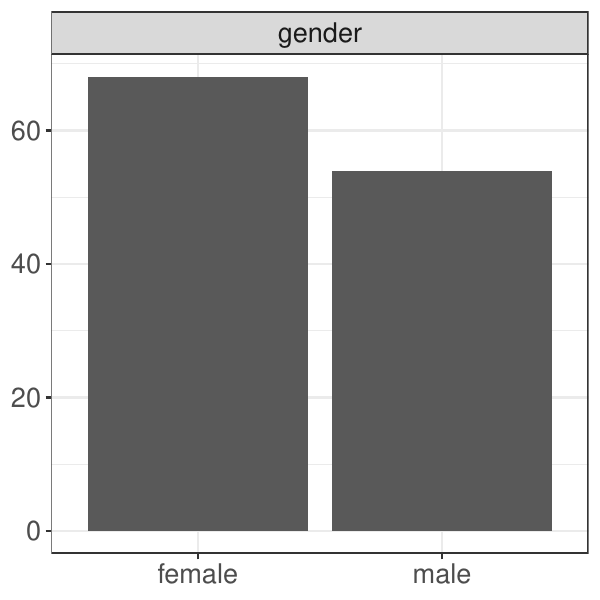}
	\includegraphics[width=0.24\linewidth]{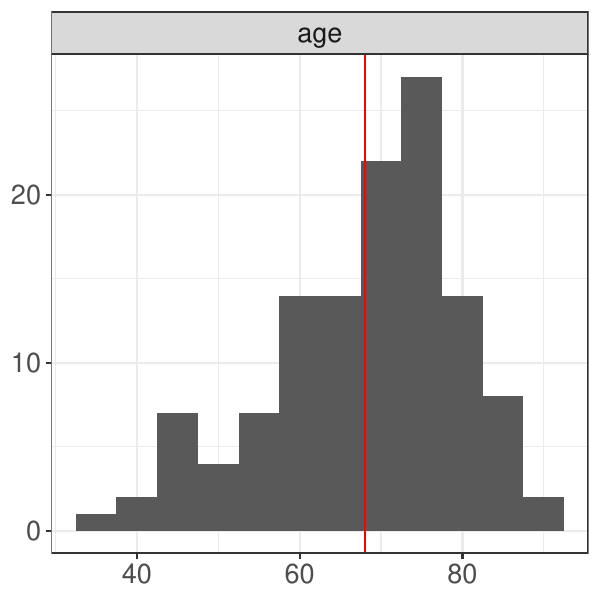}
    \includegraphics[width=0.24\linewidth]{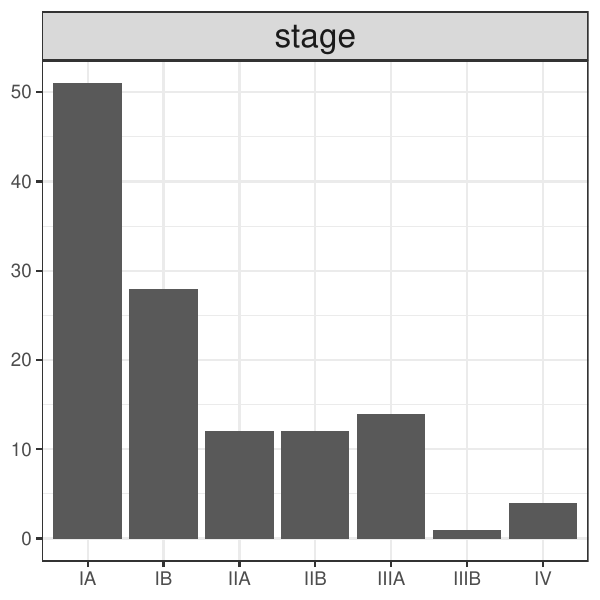}
	\includegraphics[width=0.24\linewidth]{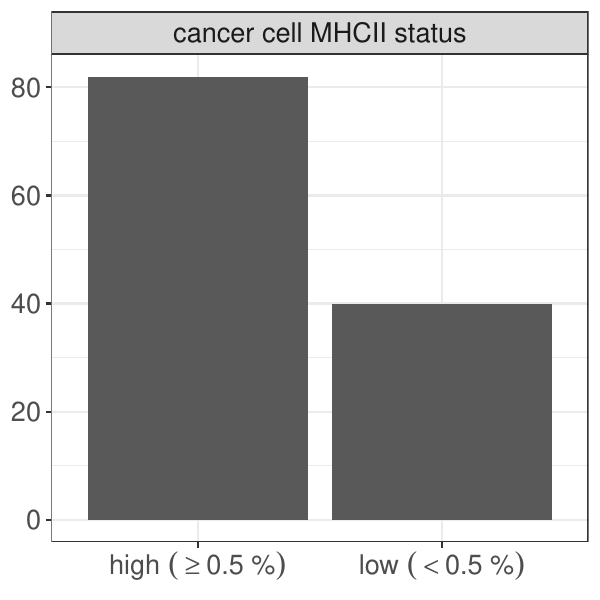}\\
    \includegraphics[width=0.24\linewidth]{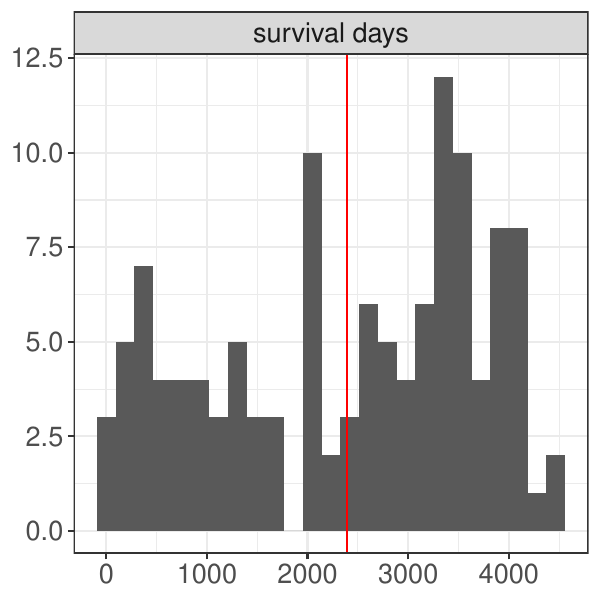}
	\includegraphics[width=0.24\linewidth]{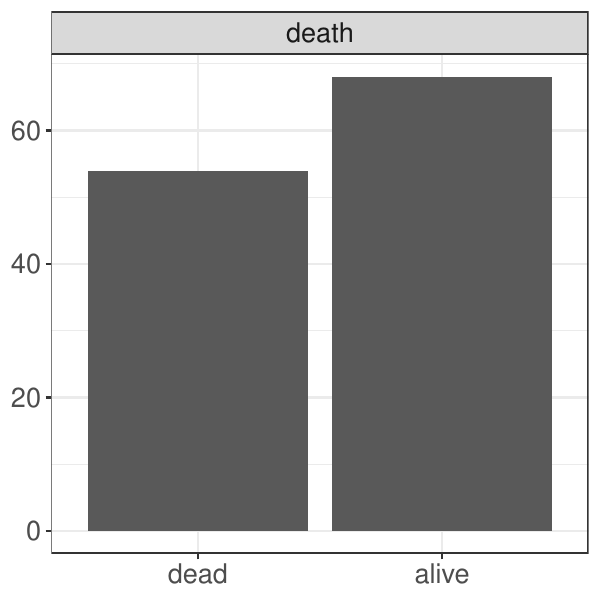}
    \includegraphics[width=0.24\linewidth]{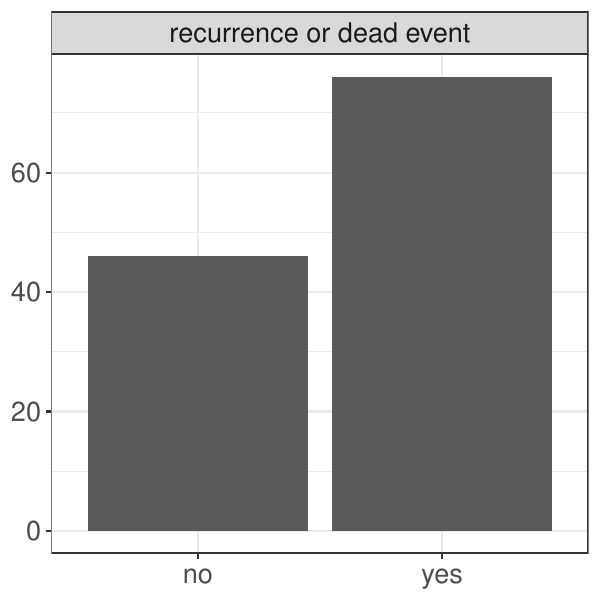}
	\includegraphics[width=0.24\linewidth]{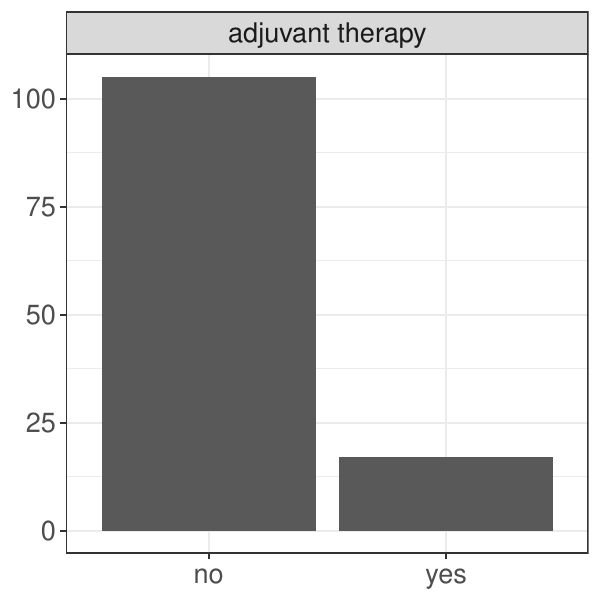}
	\caption{\label{fig:descriptive1} Summarised patients characteristics. Red lines correspond to averages.}
\end{figure}

MHCII status indicates the presence of MHCII molecules, a class of major histocompatibility complex (MHC) very important in initiating immune responses \citep[see, e.g., \ ][]{jhonson2021cancercellspecific}. The next factor is survival days, the time in days from the date of diagnosis to the date of death or the date when data collection stopped (censoring). Death factor establishes whether the participant passed away during the data collection. Recurrence or dead event informs about whether or not the participant had a recurrence or died. Finally, adjuvant therapy establishes whether or not the participant received adjuvant therapy, understood as any additional cancer treatment given after the primary. Figure \ref{fig:descriptive1} summarises this clinical information.

\section{Gibbs models}\label{sec:gibbsmodels}
Interaction is a fundamental concept between points that often cannot be observed through second-order descriptors as these functions measure correlation and not causal interaction \citep[][]{diggle2013book, baddeley2015spatialR}. Gibbs point processes models, also called Markov point processes, explicitly hypothesise that interactions occur between points in the process. These models can mimic a wide range of point patterns and can easily combine repulsion and attraction at different scales. In practice, Gibbs models only produce weak inhibition or clustering; they are helpful for modelling non-strongly clustered patterns. These models can be built based on the concept of {\it Papangelou conditional intensity} \citep{moller2004,baddeley2015spatialR}.

\subsection{Papangelou conditional intensity}
The conditional intensity function is a valuable statistic for studying and modelling point patterns \citep[see, e.g., \ ][and references therein]{diggle2021conditionalintensity}. It describes the probability of observing a cell of type $m\in \mathscr{M}$ conditional on the configuration of its neighbours in the observation window $W$. We consider a realisation of a marked (multitype) point process $\mathbf{X}$, as a set $X=\{(\mathbf{x}_i, m_i)\}_{i=1}^n$, where $\mathbf{x}_i\in W$ are the spatial locations, and $m_i\in\{1,2,\ldots, M\}$ are the types. For simplicity, we may denote a marked location $(\mathbf{x}_i,u_i)$ as $\xi_i$. If $f(X)$ is the joint probability density function of a multitype point pattern $X$, this function can be written as \citep{baddeley2015spatialR},
$$
f(X)= \zeta \left[\prod_{i=1} ^{n} B_{m_i}(\mathbf{x}_i)\right]\left[ \prod_{i<j} \Phi_{m_i,m_j}(\mathbf{x}_i,\mathbf{x}_j)\right],
$$
where $\zeta$ is known as {\it normalising constant}, and it is usually intractable \citep{moller2004}, $B_{m}(\mathbf{x})$ is a non-negative {\it first-order trend} of points of type $m$, and $ \Phi_{m,m'}(\mathbf{x},\mathbf{x}'), m,m'\in \{1,\ldots,M\}$ are pairwise interaction functions for points of types $m$ and $m'$. Then the Papangelou conditional intensity or simply {\it conditional intensity} at any point $\mathbf{x}$ of type $m$ of the observation window $W$ is defined as
$$
\lambda(\xi|X)= \lambda((\mathbf{x},m)|X)= \frac{f(X \cup \{ (\mathbf{x},m)\} )}{f(X )} = B_m(\mathbf{x})\prod_{i=1}^{n}\Phi_{m,m_i}(\mathbf{x},\mathbf{x}_i),
$$
where $X \cup \{ (\mathbf{x},m)\}$ is the extended point pattern obtained adding $(\mathbf{x},m)=\xi$ to the coordinates set. The conditional intensity at a data point $\xi_i$ is defined as $\lambda(\xi_i|X):= \lambda(\xi_i|X \backslash \xi_i)$, i.e., removing the data point $\xi_i$ in the denominator.

\subsection{Potential energy}
We can specify a Gibbs model by writing a formula for the probability density $f(X)$ as a product of the terms associated with each interaction. We call the (negative) potential of the model to the logarithm of the probability density $V(X) = \log f(X)$. The potential might be written as 
$
V(X) = V_0 + V_1(X) + V_{[\geq 2]}(X),
$
where $V_0$ is a constant and $V_1$ and $V_{[\geq 2]}$ represent the {\it spatial trend} and the {\it spatial interactions} (of all orders), respectively. This enables us to define, for instance, a pairwise interaction model; we would define the log trend $V_1(\cdot)$ and the pair potential $V2(\cdot,\cdot)$ for all points.

There is a significant technical piece behind this theory \cite[see, e.g.,\  ][pp. 126]{daley2003}. Still, the key point is that we may formulate Gibbs models with an arbitrary first-order spatial trend term $V_1(\{ \xi \}) = Z(\xi )$. Then, we have to choose the interaction term $V_{[\geq 2]}(\cdot)$ from a list of well-studied higher-order potentials whose integrability properties are known. For the conditional intensity, we have
$
\lambda(\xi |X)=\exp \left\{\Delta_{\xi } V(X) \right\},
$
where $\Delta_{\xi} V(X)=V(X \cup\{\xi \})- V(X)$.

\subsection{Fitting Gibbs models}
The idea is to model the conditional intensity in the following way,
\begin{equation}\label{eq:loglinearformUNI}
    \log \lambda(\xi |X) = B(\xi ) + \eta^{\top} Z(\xi ) + \phi^{\top}T(\xi|X),
\end{equation}
where $B(\xi)$ is an optional, real-valued function representing a baseline or offset. The first-order term $Z(\xi )$ describes spatial inhomogeneity or covariate effects. The higher-order term $T(\xi |X)$ describes interpoint interaction, and $\eta$ and $\phi$ are parameters.

\section{Fiksel interactions}\label{sec:fikselinteraction}
Motivated by some exponential decay observed in cellular contexts \citep[see, e.g.,][]{segalstephany1984cellinteractions, hui2007cellcellinteraction, li2019bayesianinteraction}, we assume that the interaction energy between every pair of immune cells (the pair potential) decreases exponentially. \cite{fiksel1986pairwiseants} proposed a bivariate pair potential function. As a step further, we display its straightforward multitype version given by 
{\footnotesize
\begin{equation}\label{eq:phifunction}
    \Phi_{ij}(r):= 
\begin{cases}
      -\infty                                       & 0\leq r < h_{ij},\\
      c_{ij}\cdot \exp{\{-\gamma_{ij} \cdot r\}}    & h_{ij} \leq r < R_{ij},\\
      0                                             & R_{ij} \leq r, 
\end{cases}
\end{equation}
}
$\{h_{ij}\}$, $\{c_{ij}\}$, $\{\gamma_{ij}\}$ and $\{R_{ij}\}$ are parameters, and $i,j\in \mathscr{M}$. The parameter $h_{ij}$ is the {\it hardcore distance} between types; points of types $i$ and $j$ must be separated at least by a distance $h_{ij}$. The {\it interaction strength} parameter $c_{ij}$ controls the type of interaction; it is zero for independent processes, positive for attractive processes and negative for repulsive processes. The {\it rate} or {\it slope} $\gamma_{ij}$ controls the decay of the interaction between $i$ and $j$ as the distance increases. The {\it interaction range} $R_{ij}$ means that points beyond this distance do not interact. 

\subsection{Inference}
\subsubsection{Pseudolikelihood}
For a multitype pairwise interaction process, the pseudolikelihood can be written as \citep{jensen1991pseudolikelihood, goulard1996parameter}
$$
\text{PL} = \left[\prod_{i=1} ^{n} B_{m_i}(\mathbf{x}_i)\right]\left[ \prod_{i<j} \Phi_{m_i,m_j}(\mathbf{x}_i,\mathbf{x}_j)\right] \cdot \exp{\left\{- \sum_{m\in \mathscr{M}} \int_W B_m(\mathbf{u})\prod_{i=1}^{n}\Phi_{m,m_i}(\mathbf{u},\mathbf{x}_i) \de \mathbf{u} \right\}},
$$
 where $\mathscr{M}$ is the set of types.

\subsubsection{Berman and Turner's device}\label{sec:bermanturner}
A Berman-Turner device is a computational tool for approximating maximum pseudolikelihood estimates \citep{bermanturner1992, baddeley2000}. It generates a set of marked points, including both a set of dummy points and the data points and forms a good {\it quadrature rule} for $W \times \mathscr{M}$. A quadrature rule is an approximation of an integral $\int_W f(\mathbf{u}) \de \mathbf{u}$ as a weighted sum $\sum_j w_j f(\mathbf{u}_j)$ of the function values at specified points (quadrature points) within the integration domain. The weights $w_j$ are {\it quadrature weights} that sum to $|W|$. \cite{baddeley2000} proposed a practical scheme to select the weights; it partitions $W$ into tiles of an equal area where each tile has a dummy point. 

Consider the Cartesian product of a set of quadrature points in $W$ and the set $\mathscr{M}$. We write the marked points as $(\mathbf{u}_j, v_{\ell})$ for $j=1,\ldots,J$ and $\ell =1,\ldots,L$ where $\mathbf{u}_j \in W$ and $k_{\ell}\in \mathscr{M}$. Then we define the indicator $z_{j \ell}$ to equal one if $(\mathbf{u}_j,k_{\ell})$ is a data point and zero if it is a dummy point. Let $w_{j \ell}$ be the corresponding weights for a linear quadrature rule in $W\times \mathscr{M}$. Then the pseudolikelihood is approximated by
\begin{equation}\label{eq:logPLbermanturner}
\log{\text{PL}} \approx \sum_{\ell = 1}^L \sum_{j=1}^J (\upsilon_{j\ell} \log \lambda_{j \ell} - \lambda_{j \ell}) w_{j \ell},
\end{equation}
where $\lambda_{j \ell}:= \lambda((\mathbf{u}_j,k_{\ell})|X)$, $\upsilon_{j \ell}:=z_{j \ell} / w_{j \ell}$, $z_{j \ell}:= 1$ if $\mathbf{u}_j$ is a data point, $z_{j \ell}:= 0$ if $\mathbf{u}_j$ is a dummy point
and the weights $w_{j \ell}$ are the areas of the tiles. The log-likelihood given in \eqref{eq:logPLbermanturner} has the form of a weighted $(w_{j \ell})$ log-likelihood of Poisson random variables $\mathcal{Y}_{j \ell}$ with expected values $\lambda_{j \ell}$ \citep{baddeley2000}. It can be handled computationally using generalised linear models techniques \citep{mccullagh1989generalized} or even generalised additive models \citep{hastie1990additivemodels}.

\subsection{Replicated point patterns methodology}
Assume that we have $g$ experimental units and that the response from unit $k\leq g$ is a  multitype point pattern $X^{k}$ observed in a window $W^{k}$. We assume that the point patterns $\left\{ X^{k} \right\}_{k=1}^g$ are independent conditional on the covariates and random effects. The conditional intensity for the $k$th point pattern is 
\begin{equation}\label{eq:modelreplication}
    \lambda^{k}(\xi |X)=\exp \left\{\Delta_{\xi} B^k(X) + \theta^{\top} \Delta_{\xi} Y^{k}(X) 
\right\},
\end{equation}
where 
$Y^k(X) := \left(Y_1^k(X), \ldots, Y_p^k(X) \right),$
are vector-valued functions representing fixed effects. Notice that random effects can be included easily as an additional term in the right of Eq. \eqref{eq:modelreplication}. Furthermore, notice that every function $f(X)$ of a point pattern $X$ can be expressed as $f(X)=f_{[1]}(X) + f_{[\geq 2]}(X)$, where 
$
f_{[1]}(X)=\sum_{\xi_i\in X} f(\{\xi_i\}),
$
is the first-order component and $f_{[\geq 2]}(X)=f(X) - f_{[1]}(X)$ is the interaction term \citep{baddeley2015spatialR}. When we apply this decomposition to the functions $B^k(X)$ and $Y^k(X)$, we retrieve the first-order components $B_{[1]}^k(\xi)$, $Y_{[1]}^k(\xi)$ that resemble the offset and the covariate effects in Eq.\eqref{eq:loglinearformUNI}, and the interaction components $B_{[\geq 2]}^k(\xi)$ and $Y_{[\geq 2]}^k(\xi)$. In our particular case, we assume that the interaction has canonical parameters; therefore, $B_{[\geq 2]}^k(\xi)$ vanishes and $Y^k_{[\geq 2]}(X)$ comes from a pairwise interaction Fiksel form (see Section \ref{sec:fikselinteraction}). 

The log-pseudolikelihood is given by
\begin{equation*}
    \log \text{PL} = \sum_{k=1}^g \sum_{(\mathbf{u}_i,m_i) \in X^k} \lambda^k((\mathbf{u}_i,m_i)|X^k) - \sum_{k=1}^g \sum_{m\in \mathscr{M}}\int_{W^k} \exp{\{ \lambda^k((\mathbf{u},m)| X^k) \}} \de \mathbf{u},
\end{equation*}
this expression is equivalent to the pseudolikelihood of a Gibbs process on the disjoint union of the windows \citep{bell2004mixedmodelreplicated, baddeley2015spatialR}.

\section{Immune cells model}\label{sec:application}
In this section, we develop a multitype Fiksel interaction model for lung cancer patients' tumour immune microenvironments. First, we define a common window for all the observations of different patients. We then combine two techniques, maximum profile pseudolikelihood and maximum pseudolikelihood, to estimate the model parameters. We then proceed to propose a set of extra models in order to compare the performance of our model with those of others. We make the comparison using residual measures and the RMSE (or its analogue in this context) that we define to be able to summarise the residuals.

\subsection{Observation window and edge correction}
Since the tissue block extraction process was consistently done on \SI{5}{\micro\metre} slides, the observation windows are the same in theory but slightly different in practice due to measurement errors and precision. To alleviate this effect, we will consider each patient's observation window $W^{\ell}$ as a dilation of the convex hull that contains the data (by $1/\sqrt{1 - \omega_{\ell} / n_{\ell}}$; $n_{\ell}$ is the number points of $X^{\ell}$, and $\omega_{\ell}$ is the number of vertices of the convex hull of the patient $\ell$) \citep{ripley1977ripras}. Once we have these windows, the final observation window, which is also common for all patients, is defined as
\begin{equation}\label{windowsintersection}
    W:=\bigcap_{\ell = 1}^{151} W^{\ell}.
\end{equation}

When the goal is to make inference, it is important to assume that the data is a realisation of a finite point process defined only within $W$ (bounded case) or a partially observed realisation of a point process that extends along a bigger domain only through the window $W$ (unbounded case). In our context, we must assume that our point patterns are partially observed realisations. This is due to two reasons, first is that the tissues analysed before imaging are just samples of larger tissue (lungs in this case). Second, we cut the windows to make a common window through Eq. \eqref{windowsintersection}. 

There can be {\it edge-effect} problems in the unbounded case \cite{baddeley2000} since some information might come from unobserved points outside the final observation window. There are several methods in the literature to alleviate this type of effect \citep[see, e.g., \ ][and references therein]{ripley1988, baddeley2000, chiuetal2013stochastic}. In our case, we use the well-known {\it border method} \citep{ripley1988}, which obtains the pseudo-likelihood integration domain by cutting a width margin $r$ from the original observation window.

\subsection{Trend and interaction terms}
\begin{figure}[h!tb]
	\centering
	\includegraphics[height=1.568in]{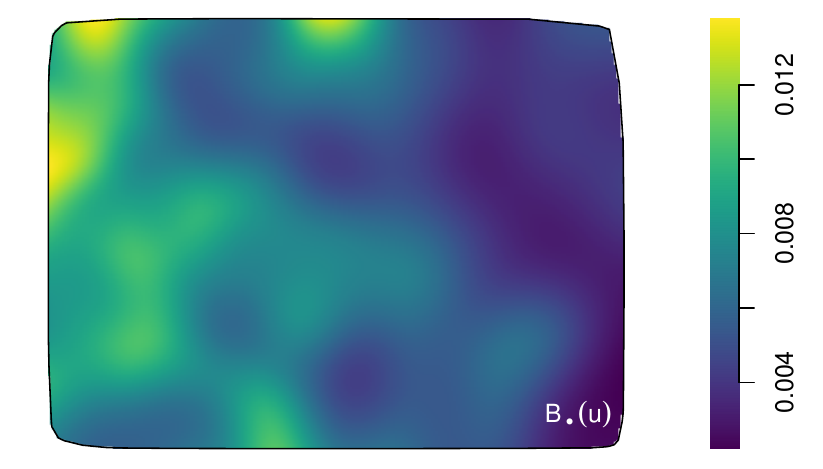}
	\includegraphics[height=1.568in]{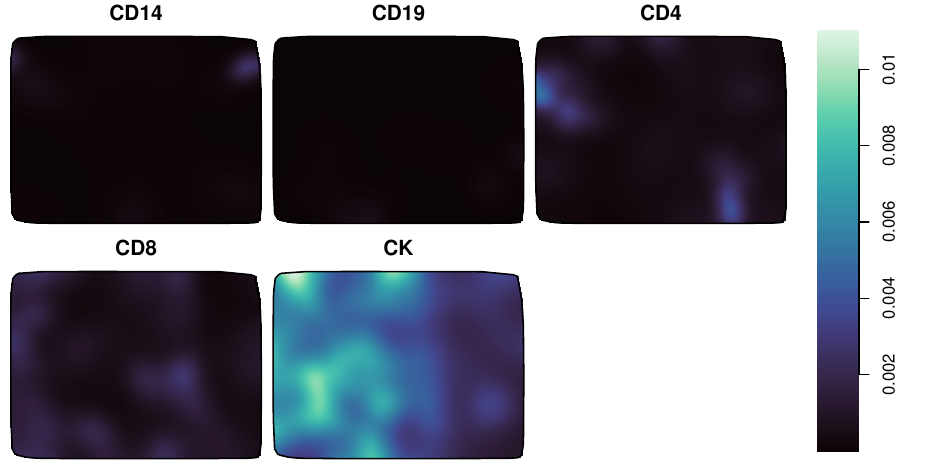}
	\caption{\label{fig:firstorderintensities} Left: Non-parametric estimate of the overall first-order intensity function, including all the cell types. Right: First-order intensity functions by cell type.}
\end{figure}

We incorporate inhomogeneity into our model through an offset in which we non-parametrically estimate the total first-order intensity of each of our point patterns. The total intensity function is defined as 
$
B_{\bullet}(\mathbf{u})=\sum_{m\in \mathscr{M}}B_m(\mathbf{u}),
$
\citep{baddeley2015spatialR}. In this way, we generate a smooth estimate of the expected value of the number of immune cells at each point in the observation region, considering all cell types simultaneously. This estimation is made through a spatial Gaussian kernel with adaptive bandwidth \citep[see, e.g.\ ][and references therein]{Davies2018kernel}. This estimator is defined as follows,
\begin{equation*}
\hat{B}_{m}\left(\mathbf{u} \right) =\frac{1}{e_{\epsilon} \left(\mathbf{u}\right) }\sum_{\mathbf{u}_i \in X_m}{K_{\epsilon(\mathbf{u}_i)}\left(\mathbf{u}-\mathbf{u}_{i}\right)}, \qquad \mathbf{u} \in W, m\in \mathscr{M},
\end{equation*}
where $K(\cdot)$ is a Gaussian kernel, $\epsilon(\cdot)$ is a bandwidth function and
$
e_{\epsilon} \left(\mathbf{u} \right)
$
is an edge correction \citep{marshalhazelton2010boundarieskernel}.  The estimates for $B_{\bullet}(\mathbf{u})$ and for $B_{\text{CD14}^+}(\mathbf{u})$, $B_{\text{CD19}^+}(\mathbf{u})$, $B_{\text{CD4}^+}(\mathbf{u})$, $B_{\text{CD8}^+}(\mathbf{u})$ and $B_{\text{CK}^+}(\mathbf{u})$ are shown in Figure \ref{fig:firstorderintensities}.

Additionally, we consider the design covariates, which, although not spatial, correspond to factors that influence the overall conditional intensity. These factors have been explained in Section \ref{sec:data} and are related to patients' clinical information; we denote them by $Z$.

We introduce the interaction between cells into the model through the term $T(\mathbf{u},m)$. This inclusion of interaction entails the assumption of several facts; in this case, we assume that we have the same interaction for all patients, that is, the Fiksel interaction defined by the $\Phi_{ij}(r)$ function given in Eq. \eqref{eq:phifunction}. We also assume we have the same sets of parameters $\{h_{ij}\}$, $\{\gamma_{ij}\}$,$\{R_{ij}\}$ and $\{c_{ij}\}$ for all patients. Modifying this assumption would correspond to having a previously identified mechanism that could alter these parameters for different groups of patients or, in the worst case, assuming that each patient has their own isolated set of parameters. This decision would make the model highly complex and likely cause overfitting problems. Therefore, we opt to choose the most parsimonious model in this case.

\subsubsection{Irregular parameters}
When a parameter of a point process model does not appear in the log-linear form \eqref{eq:loglinearformUNI}, it is called {\it irregular} \citep{baddeley2000, baddeley2015spatialR}. In contrast, the other parameters are called {\it regular}. Consider, for example, a model of the form,
$$
\log \lambda_{\vartheta}((\mathbf{u},m)|X) = \varphi^{\top} \cdot Z((\mathbf{u},m), \psi|X)
$$
where $\psi$ is the vector of irregular parameters, $\varphi$ is the vector of regular ones and $\vartheta := (\varphi, \psi)$. For every fixed value of the irregular parameters, the model is log-linear in the regular ones, i.e., if we fix the values of $\psi$, then the model is log-linear in $\varphi$; this model can be fitted using maximum pseudolikelihood over $\varphi$ \citep{besag1975pseudolikelihood}. For retrieving a maximum {\it profile pseudolikelihood} estimate, we assign a value to $\psi$, then, the pseudolikelihood $\text{PL}(\varphi, \psi)$ can be maximised over all possible values of $\varphi$,
$
\text{PPL}(\psi)=\max_{\varphi}\text{PL}(\varphi, \psi).
$
The maximum pseudolikelihood estimate of $\vartheta$ can be obtained by maximising the profile pseudolikelihood over $\psi$. 

{\it Hardcore distances}

A maximum likelihood estimator of the hardcore radii is the minimum nearest-neighbour distance amongst the points with different labels \citep[see, e.g., \ ][]{baddeley2015spatialR}. Given that we have replicated patterns, we choose the minimum across the replicates for every matrix entry, so distinct points are not permitted to come closer than this minimum apart. The estimate is given by 
{\footnotesize
$$
\hat{h}_{ij} = 
\begin{pmatrix} 
            &\text{\color{gray} CD14}    &\text{\color{gray} CD19}   &\text{\color{gray} CD4}    &\text{\color{gray} CD8}    &\text{\color{gray} CK} \\
\text{\color{gray} CD14} &0.498   &0.496  &0.497  &0.497  &0.495 \\
\text{\color{gray} CD19} &0.496   &0.499  &0.496  &0.495  &0.481  \\
\text{\color{gray} CD4}  &0.497   &0.496  &0.498  &0.496  &0.496  \\
\text{\color{gray} CD8}  &0.497   &0.495  &0.496  &0.498  &0.497  \\
\text{\color{gray} CK}   &0.495   &0.481  &0.496  &0.497  &0.499
\end{pmatrix}.
$$
}
We observe similar values in the $\hat{h}_{ij}$ entries. This means that the tumour immune microenvironment cells could share a common hardcore distance, which could simplify the model since instead of considering a matrix of hardcore distances, we could only consider a positive scalar, given, for example, by $\min_{ij}\{\hat{h}_{ij}\}=\SI{0.481}{\micro\metre}$. Models with this type of simplification are shown in Section \ref{sec:severalmodels}.

{\it Interaction range and rate or slope}

\begin{figure}[h!tb]
	\centering
	\includegraphics[width = 0.8\linewidth]{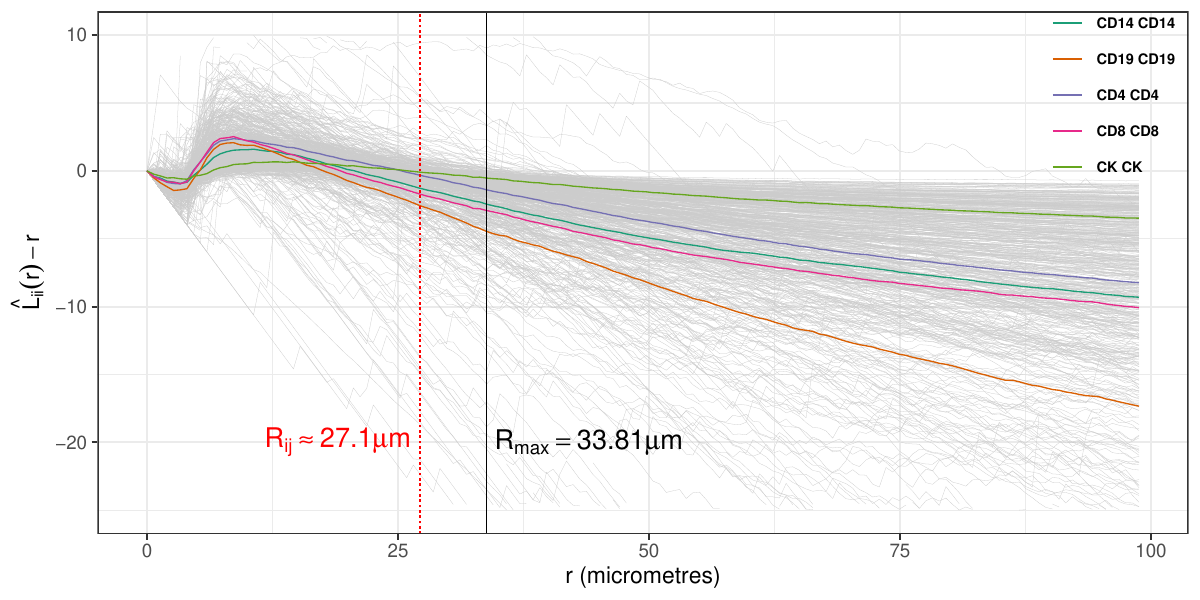}
	\caption{\label{fig:Lfunctions} Inhomogeneous $L$-functions of every patient for cells of the same type (grey lines). The coloured lines correspond to the averages per cell type. The vertical black line represents the maximum value assigned to the interaction range $R_{ij}$. The red line corresponds to the final interaction range chosen by maximising the profile pseudolikelihood.}
\end{figure}

We must provide a suitable range of values for the parameters to apply the maximisation over the profile pseudolikelihood. For the interaction range, we can revise Ripley's $K$-function $K_{ij}(r)$ (of its variance stabilised version, the $L$-function, $L_{ij}(r)$) in its multitype inhomogeneous version \citep{moller2004}. Roughly speaking, this function represents the expected value of the count of events of the type $j$, weighted by the reciprocal of the intensity at each point of type $j$, within distance $r$ of an arbitrary event of type $i$. It could be estimated by 
$$
\hat{K}_{ij}(r)=\frac{1}{|W|} \sum_{\mathbf{u}_{\ell}\in X_i} \sum_{\mathbf{u}_{k}\in X_{j}} \frac{\mathbf{1} \{||\mathbf{u}_{\ell}-\mathbf{u}_k|| \leq r\}}{\hat{B}_i(\mathbf{u}_{\ell})\hat{B}_j(\mathbf{u}_k)}e(\mathbf{u}_{\ell}, \mathbf{u}_k;r), \quad i,j\in \mathscr{M},
$$
where $\mathbf{1}\{\cdot \}$ is the indicator function, and $e(\cdot)$ is an edge correction \citep{baddeley2015spatialR}. The $L$-function is intended to stabilise the $K$-function variance and it is defined as $L_{ij}(r):=\sqrt{K(r)/ \pi}$. When the points. For illustration purposes, Figure \ref{fig:Lfunctions} displays the $L$-functions of the cells of the same type.
$L$-functions of the same type $(L_{ii}(r))$ are the usual $L$-functions of the process $X_i$ of points of the type $i$, meaning that the same interpretation applies as if there was no labelling. For example, the typical benchmark of $L(r)=r$ for Poisson processes still applies here. We select a maximum possible interaction range so we may see a stable behaviour of the mean $L$-function (understood as the classical functional mean) across the patients behind such selected maximum; this value is $\SI{33.81}{\micro\metre}$ and it is displayed as a vertical black line in Figure \ref{fig:Lfunctions}.

For a range for the rate or slope, we decide to take into account the scale of the data and assign $\gamma_{ij} \in [0.2, 0.2]\SI{}{\micro\metre}$. After applying the procedure of maximising the profile pseudolikelihood, we retrieve the estimations for $R_{ij}$ and $\gamma_{ij}$
{\footnotesize
$$
\hat{R}_{ij} = 
\begin{pmatrix} 
        27.11   &21.00  &18.03  &20.03  &24.42 \\
        21.00   &27.11  &19.08  &19.32  &25.92 \\
        18.03   &19.08  &27.11  &16.40  &23.13 \\
        20.03   &19.32  &16.40  &27.11  &24.55 \\
        24.42   &25.92  &23.13  &24.55  &27.21 \\
\end{pmatrix},
\qquad
\hat{\gamma}_{ij} = 
\begin{pmatrix} 
        0.110       &-0.066     &-0.031     &-0.041     &-0.082 \\
        -0.066      &0.073      &-0.071     &-0.080     &-0.077 \\  
        -0.031      &-0.071     &0.111      &-0.052     &-0.048 \\
        -0.041      &-0.080     &-0.052     &0.111      &-0.043 \\
        -0.082      &-0.077     &-0.048     &-0.043     &0.200 \\
\end{pmatrix}.
$$
}

\subsubsection{Regular parameters}\label{sec:regularparameters}
There are ten regular parameters $\nu_1,\ldots,\nu_X$ and $c_{ij}$, i.e., all terms that appear in the conditional intensity log-linear form, one of them is the intercept, eight of them  are the coefficients for each clinical covariates. The estimation procedure is done through the pseudolikelihood and the Berman-Turner approximation, considering the replicates. The estimated coefficients are shown in Table \ref{tab:coefficients}.
\begin{table}[h!tb]
\centering
\fbox{%
\footnotesize
\begin{tabular}{lllll} 
{\bf Coefficients}       &Estimate $(\eta)$      &$\exp\{ \eta \}$& Std. Error & $p$-value\\ \hline
(Intercept)              &$-5.459$              &$0.0043$    &$8.69\times 10^{-3}$  &$< 2\times 10^{-16}$ ***\\
Gender (Masculine)       &$-4.40\times 10^{-2}$ &$0.9570$    &$2.38\times 10^{-3}$  &$< 2\times 10^{-16}$ ***\\
Age at diagnosis         &$-4.70\times 10^{-3}$ &$0.9953$    &$1.07\times 10^{-4}$  &$< 2\times 10^{-16}$ ***\\
Stage &&&&\\
\hspace{0.5cm}    IB     &$2.50\times 10^{-2}$  &$1.0253$    &$3.12\times 10^{-3}$  &$1 \times 10^{-15}$ ***\\
\hspace{0.5cm}    IIA    &$9.00\times 10^{-2}$  &$1.0942$    &$4.17\times 10^{-3}$  &$< 2\times 10^{-16}$ ***\\
\hspace{0.5cm}    IIB    &$-9.92\times 10^{-2}$ &$0.9056$    &$4.05\times 10^{-3}$  &$< 2\times 10^{-16}$ ***\\
\hspace{0.5cm}    IIIA   &$-1.00\times 10^{-1}$ &$0.9047$    &$4.13\times 10^{-3}$  &$< 2\times 10^{-16}$ ***\\
\hspace{0.5cm}    IIIB   &$-7.73\times 10^{-2}$ &$0.9256$    &$1.39\times 10^{-2}$  &$2.92 \times 10^{-8}$ ***\\
\hspace{0.5cm}    IV     &$1.15\times 10^{-1}$  &$1.1213$    &$6.37\times 10^{-3}$  &$< 2\times 10^{-16}$ ***\\
MHCII status (low $<0.5\%$)  &$-8.30\times 10^{-2}$ &$0.9204$ &$2.49\times 10^{-3}$  &$< 2\times 10^{-16}$ ***\\
Survival days            &$-1.16\times 10^{-5}$ &$1.0000$    &$9.38\times 10^{-7}$  &$< 2\times 10^{-16}$ ***\\
Death (Yes)              &$-3.60\times 10^{-2}$ &$0.9647$    &$5.06\times 10^{-3}$  &$1.22 \times 10^{-12}$ ***\\
Recurrence (Yes)         &$-5.52\times 10^{-4}$ &$0.9995$    &$4.97\times 10^{-3}$  &$0.911$   \\
Adjuvant therapy (Yes)   &$-3.99\times 10^{-2}$ &$0.9609$    &$3.52\times 10^{-3}$  &$< 2\times 10^{-16}$ ***\\
\end{tabular}}
\caption{\label{tab:coefficients} Estimated regression coefficients and $p$-values in the fitted model for the conditional intensity of the immune cells.}
\end{table}
All model design covariates were statistically associated with conditional intensity except for the recurrence variable ($p$-value of 0.911); i.e., the factor that reports whether the patient had a recurrence or died has no statistical impact on conditional intensity. The values in Table \ref{tab:coefficients} come from a generalised linear model, as detailed in Section \ref{sec:bermanturner}. Therefore, it should be noted that the $p$-values are calculated based on traditional mechanisms. This means that the significance depends on the number of observations, roughly seven million in this case. With such a large number of observations, it is logical and expected that almost all the factors become statistically significant \citep[see, e.g., \ ][]{lin2013pvalueproblem}, which is what happens in this case. To avoid a vague interpretation, we focus on the regression coefficients $(\exp\{\eta \})$. Factors associated with reduced conditional intensity are gender, where men generally have lower immune cell counts than women; a low MHCII status has less intensity than high MHCII status; and death, where those who died showed less immune cell density than those who did not. Patients who received adjuvant therapy also show lower counts than those who did not; this may occur as immune cells may be found within cancerous tissues targeted by adjuvant therapies to be removed or killed. Regarding the disease status, we can observe an intensity increase in patients in stage IV compared to those in stage IA and a decrease in patients in stage III. 

{\it Fitted interaction strength}

\begin{figure}[htb]
	\centering
	\includegraphics[width = .4\linewidth]{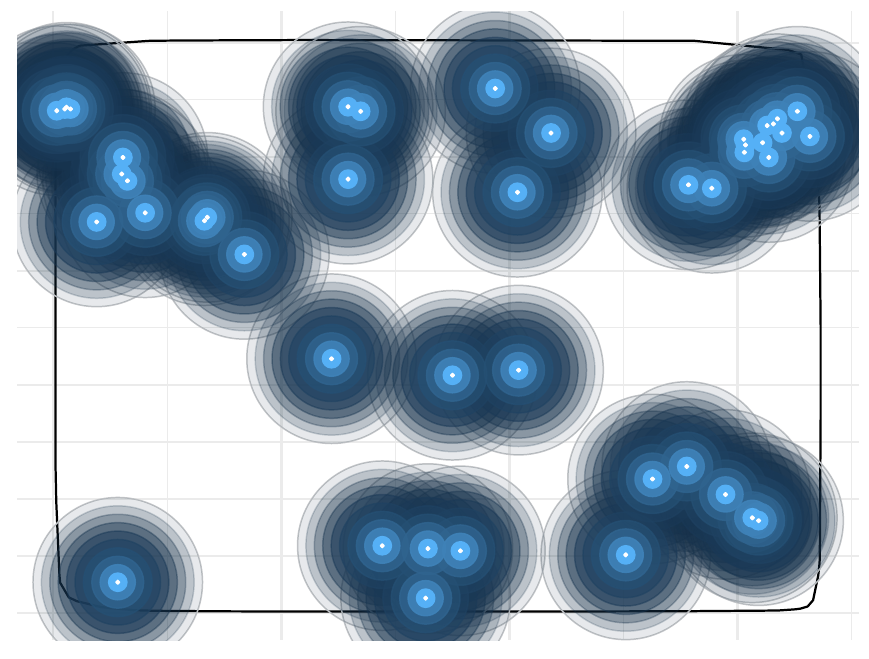}
	\includegraphics[width = .4\linewidth]{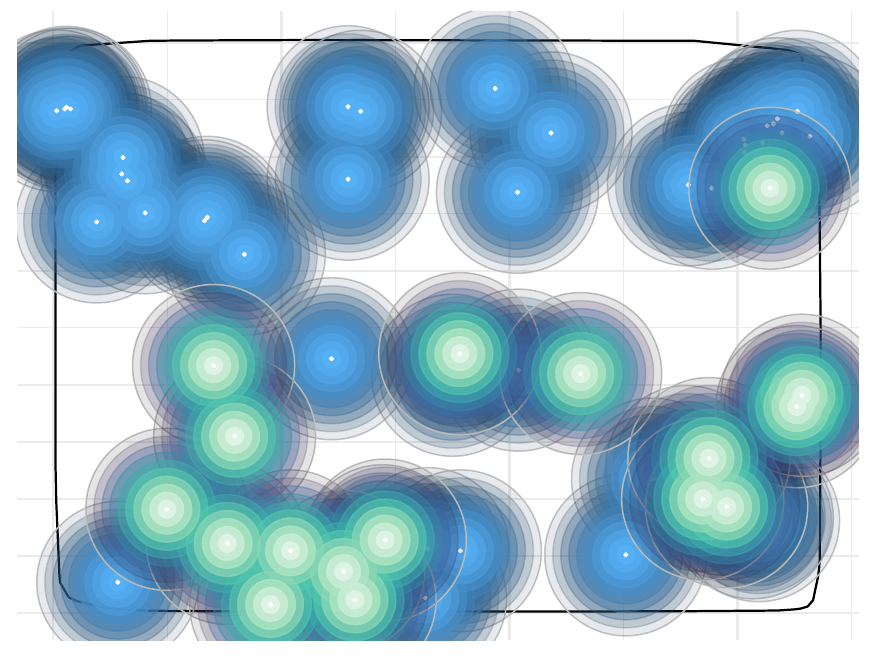}
	\caption{\label{fig:influencezones} Locations of CD14$^+$ (left) and CD14$^+$ and CD19$^+$ cells (in two different colour tones); the sizes of the circles are the chosen influence zones of the cells under the Fiksel model interaction for cells of the same type $\Phi_{11}(r)$ (left) and different type $\Phi_{12}(r)$. The colours vanish according to the strength of the interaction function $\Phi$. The two plots correspond to the very same patient.}
\end{figure}

The other parameter of the model is the strength of the Fiksel interaction term $c_{ij}$. 
{\footnotesize
$$
\hat{c}_{ij} = 
\begin{pmatrix} 
            1.3052  &0.9995 &0.9994 &0.9998 &0.9996 \\
            0.9995  &1.2171 &0.9997 &0.9996 &0.9996 \\
            0.9993  &0.9997 &1.1951 &0.9988 &0.9999 \\
            0.9998  &0.9996 &0.9988 &1.4694 &0.9993 \\
            0.9996  &0.9996 &0.9999 &0.9993 &1.0473
\end{pmatrix}.
$$
}
For illustration purposes, in Figure \ref{fig:influencezones}, we show the conditional intensities of CD14$^+$ cells considering their interaction with cells of the same type and their interaction with CD19$^+$ cells of a single patient included in our sample. 
The small size of the white dots represents the minimum distance of repulsion $\hat{h}_{ij}$, i.e., CD14$^+$ cells are prohibited from locating within \SI{0.498}{\micro\metre} of other CD14$^+$ cells and within \SI{0.499}{\micro\metre} of CD19$^+$ cells. Beyond this distance, the attraction decays exponentially with the distance according to the $\Phi_{ij}(r)$ function given in Eq. \eqref{eq:phifunction} for cells of the same type. In the case of cells of a different type, the $\Phi$ function does not decay; instead, it increases due to the sign of the $\gamma_{ij}$ parameter. However, the magnitude of this quantity is generally smaller for different cells, which makes the interaction's strength less in these cases. This type of behaviour, where the magnitudes of interaction are observed to be so small for cross-terms, makes us think of simpler alternative models, for example, an interaction model only within types. These models are discussed in Section \ref{sec:severalmodels}.

\begin{figure}[h!tb]
	\centering
	\includegraphics[width = 1\linewidth]{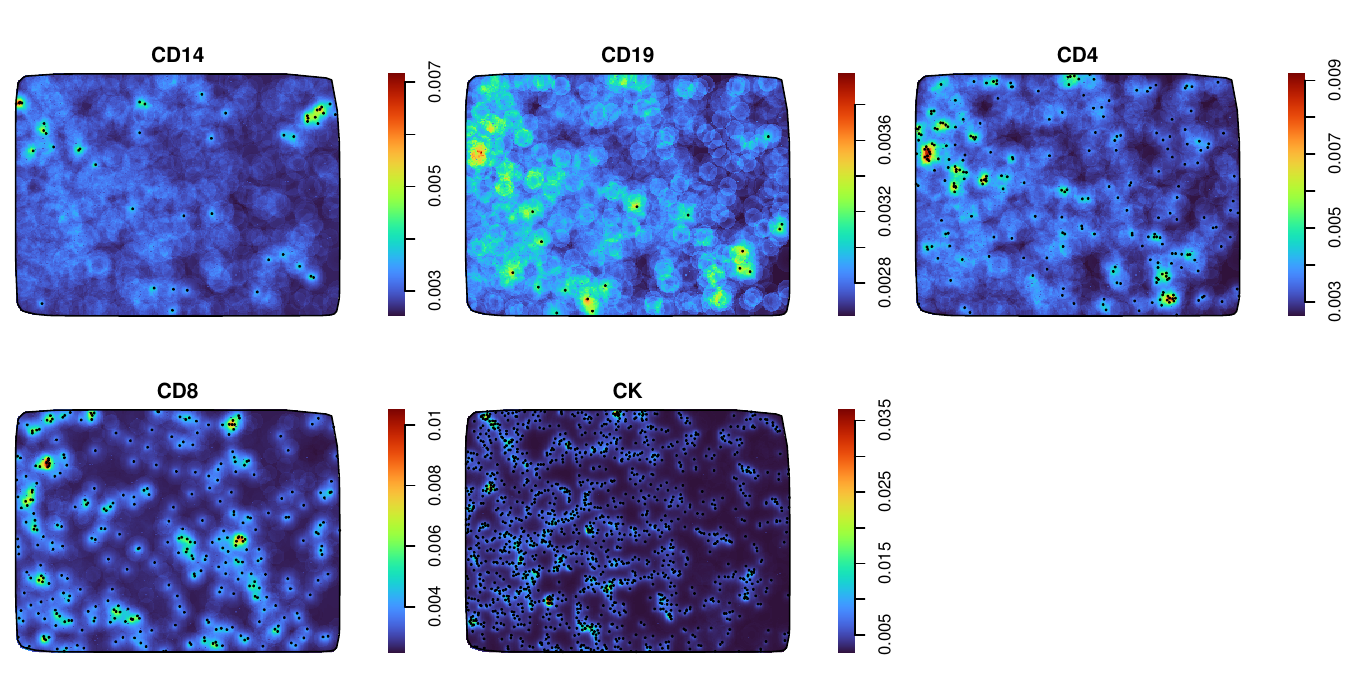}
	\caption{\label{fig:cif} Estimated conditional intensity of a patient's immune cells according to the maximum pseudolikelihood fit of the log-linear model. The superimposed black dots represent the actual locations of cells of each type.}
\end{figure}

Figure \ref{fig:cif} shows each cell type's fitted conditional intensity $\log \hat{\lambda}(\xi|X)$ from an arbitrarily chosen patient. This conditional intensity is evaluated in a regular mesh in the observation window $W$. We may see how the adjustment is satisfactory even in cases with few points, such as CD14$^+$, CD19$^+$ and CD4$^+$. We see a better fit for cell types with more points, CD8$^+$ and CK$^+$. This evaluated conditional intensity strongly suggests that the fit is adequate, considering the model is simultaneously set up for all patients. This gives us an idea of how suitable the model with the Fiksel interaction is for tumour immune microenvironment modelling.

\subsection{Assessing the model}
We want to test whether or not the model is working well; this model evaluation can be done in many ways. In this work, we compute some residual summaries of the proposed model. Residuals from point process models are a proper diagnostic measure for comparisons \citep{baddeley2015spatialR}.

\subsubsection{Comparing several models} \label{sec:severalmodels}
 We consider several models to be able to compare their performance and finally opt for one or some. To do this, we rely on the fact that our model comprises three fundamental parts: an offset, other first-order effects (design covariates), and second-order effects (Fiksel's interaction). In order to understand how good the model is, we will propose several models where these different parts are included or not. 

We then consider three sets of models. In the first one, we include four models that have in common the multitype Fiksel-type interaction function that we propose in this article. The first is the model described in Section \ref{sec:fikselinteraction}, containing all the components (Fiksel 1). The second is a model without first-order effects (Fiksel 2); the third considers only the effects of the design covariates but not the offset (Fiksel 3). Finally, the fourth model considers first-order effects, but an interaction function that, although it is Fiksel type, assumes that the interaction does not depend on different types of immune cells; that is, the interactions only occur between cells of the same type (Fiksel 4).
 
The literature on multitype Gibbs models is not very huge as far as we know. Some classical univariate interaction functions have been extended to the multitype case \citep[see, e.g., \ ][]{diggle2006_amacrinebook, grabarnik2009hierachical}. For example, multitype Strauss, Hardcore and Strauss Hardcore models \citep{baddeley2015spatialR}. For the next set of models for comparison, we opt for holding the first-order terms and changing the pairwise interaction functions. So we choose multitype Strauss (Strauss), Hardcore (Hardcore), and Strauss Hardcore (Srt Hardcore) models. The interaction pairwise functions for these models are shown in Table \ref{tab:alternativeinteractions}.
\begin{table}[h!tb]
\centering
\fbox{%
\footnotesize
\setlength{\tabcolsep}{3.5pt}
\begin{tabular}{ccc} 
{\bf Strauss model}      &{\bf Hardcore model}      &{\bf Strauss Hardcore model}\\ \hline  
\vspace{0.5mm}
$    \Phi_{ij}(r)=
\begin{cases}
      0                      & r > R_{ij},\\
      \log \gamma_{ij}    & r \leq R_{ij},
\end{cases}
$
&
$    \Phi_{ij}(r)= 
\begin{cases}
      0           & r < R_{ij},\\
      -\infty     & r \leq R_{ij},
\end{cases}
$
&
$    \Phi_{ij}(r)= 
\begin{cases}
      -\infty               & r < h_{ij},\\
      \log \gamma_{ij}     & h_{ij} \leq r \leq R_{ij},\\
      0                     & r > R_{ij}.
\end{cases}
$
\end{tabular}}
\caption{\label{tab:alternativeinteractions} Pairwise interaction functions of the alternative models.}
\end{table}

We have also decided to generate a last model with the first-order effects but no associated interaction function (Poisson). This model corresponds to a Poisson model to explain the conditional intensity function. It should be noted that the conditional intensity coincides with the first-order intensity of an inhomogeneous Poisson process under this assumption of no interaction between points \cite{baddeley2015spatialR}.

\subsubsection{Residuals} 
\cite{baddeleyetal2005residual} defined residuals and residual plots for Gibbs models for spatial point processes, providing a strategy for model criticism in spatial point process models. Their techniques resemble the existing methods for linear models, i.e., they represent the differences between the data and the fitted model. The {\it raw residual measure} can be defined as
$$
\mathscr{R}_{m}^k(B)=N(X_{m}^{k} \cap B) - \int_B \hat{\lambda}^k((\mathbf{u},m)|X^k) \de \mathbf{u}, \quad \forall B \subseteq W, m\in \mathscr{M}, k\leq g.
$$
This function can be estimated in any subset of the observation window; that is the rationale behind the term ``measure''. Usually, a regular window partition is set to estimate the measure in each pixel as per density estimations.

In practice, the residuals are often scaled to calculate, for example, standardised residuals. The analogue to Pearson's residuals in this context is given by
$$
\mathscr{R}^{\star k}_m (B)=\sum_{\mathbf{u}_i\in X_{m}} \hat{\lambda}^k ((\mathbf{u}_i,m)|X^k)^{-1/2} - \int_B \hat{\lambda}^k((\mathbf{u},m)|X^k)^{1/2} \de \mathbf{u}.
$$
There is a third version of the residual measure called {\it inverse $\lambda$ residuals}  
$$
\mathscr{R}^{\dag k}_m (B)=\sum_{\mathbf{u}_i\in X_{m}} \frac{\mathbf{1}\left\{ \hat{\lambda}((\mathbf{u}_i,m)|X^k)>0\right\}}{\hat{\lambda}^k((\mathbf{u}_i,m)|X^k)} - \int_B \mathbf{1}\left\{ \hat{\lambda}^k((\mathbf{u},m)|X^k)>0\right\} \de \mathbf{u}.
$$

\begin{figure}[htb]
	\centering
	\includegraphics[width = .85\linewidth]{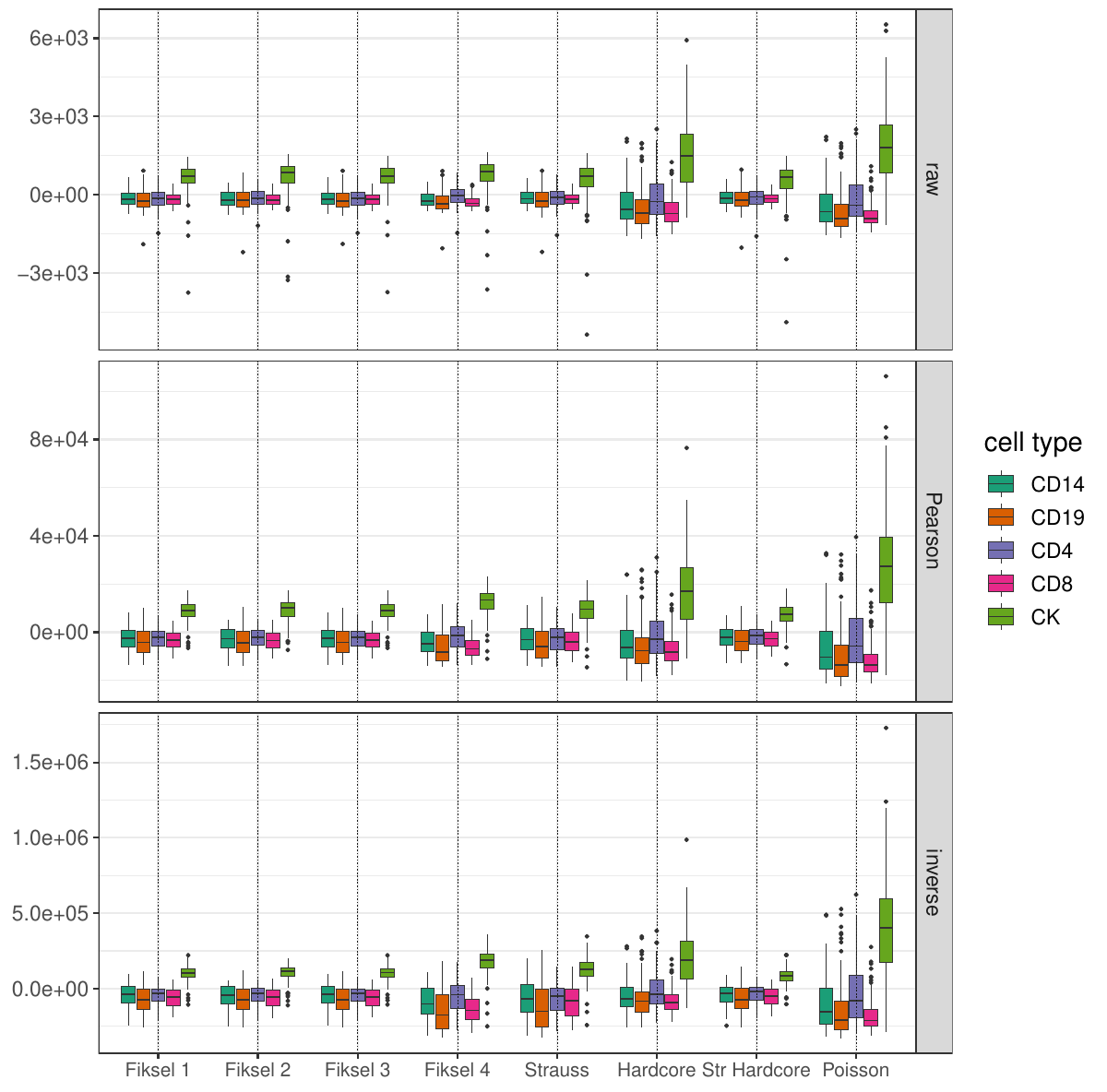}
	\caption{\label{fig:boxplotmodels} Boxplots of total raw, Pearson and inverse residuals per immune cell type against the eight models fitted.}
\end{figure}

For comparison purposes, we need to summarise some residual measure $\mathscr{R}(B), \mathscr{R}_{(\text{P})}(B)$ or $\mathscr{R}_{(\text{I})}(B)$, thus we consider the total value (the integral) of these measures over the observation window $W$. As we have five different types of cells, we may obtain a total value for each patient. Then we retrieve $122\times 5$ total residuals. Figure \ref{fig:boxplotmodels} summarises these residuals for each proposed model. 

From Figure \ref{fig:boxplotmodels}, we can glimpse several exciting things; we see how the three residuals provide roughly the same information, although on different scales. We also see how CK+ cancer cells seem the most difficult to model since their residuals are the furthest from zero in all models. The model that we have proposed and its variants, that is, those with multitype Fiksel interactions, generally present a similar and very adequate behaviour, except perhaps for the Fiksel 4 model, that is, the one where it is assumed that immune cells of different types do not interact with each other; this model has greater variability than its counterparts. This good behaviour of the models of the first set allows us to glimpse that this function is appropriate to model this type of cells, which is in harmony with our motivation (see Section \ref{sec:fikselinteraction}) to use this type of interaction.

The Poisson model is presented as the most inadequate since it is not only the one that is furthest from zero but also presents the greatest variability; this suggests that the interaction between cells must be a fundamental part of any model proposed for this type of tumour immune microenvironments. The models whose interaction functions are Strauss or Hardcore do not generate good models either. Although the residuals of these cases are closer to zero than in the Poisson case, their variability is greater than that of the other cases considered. Of the alternative interaction functions, the Strauss Hardcore is the one that best manages to model immune cells; this model is the most competitive that we can find among the multitype models that are currently known. 

We especially highlight the Fiksel 2 model since, although far from being the best, it is surprisingly good at modelling cells without having first-order information. If we wanted to simplify the model, we could do without the first-order information and still obtain a successful model. This has positive implications in practice; for example, we do not need prior knowledge of the patients' clinical conditions to obtain information on the distribution of cells in the tumour immune microenvironment. Although we strive to obtain reasonable estimates of the offset (the expected value of the counts per unit area at each point in the observation window), it does not appear critical to obtaining a good model; the model Fiksel 3 confirms this fact as well. 

{\it Root mean square error}

We wish to summarise our residuals in order to provide an overall notion of the performance of the models. For doing so, we first consider an overall residual measure across the type-cells given by $\mathscr{R}^k_{\bullet}:=\sum_{m\in \mathscr{M}}\mathscr{R}^k_{m}$. We then define the {\it Root mean square error} in this context as
\begin{equation}\label{eq:RMSE}
    \text{RMSE} = \sqrt{\frac{1}{g}\sum_{k=1}^g\left(\int_{W} \de \mathscr{R}^k_{\bullet}\right)^2}.
\end{equation}
Notice that we can straightforwardly extend this definition to Pearson's and inverse residuals. We then compute these residuals for every one of the considered alternative models. 

\begin{table}[h!tb]
\centering
\fbox{%
\footnotesize
\setlength{\tabcolsep}{3.5pt}
\begin{tabular}{llccccc}
&&&&\multicolumn{3}{c}{RMSE from residuals} \\ \cline{5-7}
\multicolumn{1}{l}{\textbf{Model }} &Interaction & Offset &Covariates &$\mathscr{R}(B)$ 
&$\mathscr{R}^{\star k}_m (B)$		
&$\mathscr{R}^{\dag k}_m (B)$\\ \hline
    {\small Fiksel 1}	   &{\small  Multi Fiksel}                &Yes   &Yes  &$1262.73$	&$16751.64$	&$282049.90$\\
    {\small Fiksel 2}	   &{\small  Multi Fiksel}                &No 	 &No   &$1395.72$  	&$17752.35$	&$291855.70$\\
    {\small Fiksel 3}	   &{\small  Multi Fiksel}                &No	 &Yes  &$1268.58$	&$16860.80$	&$283011.50$\\
    {\small Fiksel 4}	   &{\small  Multi Fiksel (no crossed)}   &Yes   &Yes  &$1102.42$	&$17269.40$	&$410336.70$\\
    {\small Strauss}	   &{\small  Multi Strauss}               &Yes   &Yes  &$1332.00$	&$20437.87$	&$443944.80$\\
    {\small Hardcore}	   &{\small  Multi Hardcore}             &Yes   &Yes  &$2971.19$	&$35312.84$	&$424544.00$\\
    {\small Str Hardcore}  &{\small  Multi Strauss Hardcore}     &Yes 	 &Yes  &$1276.67$	&$15586.93$	&$264267.50$\\
    {\small Poisson}	   &{\small  None (Poisson model)}        &Yes   &Yes  &$2039.28$	&$31471.21$	&$488105.40$\\

\end{tabular}}
\caption{\label{tab:RMSE} Estimated root mean square errors (RMSE) for the eight proposed models for immune cells. The type of interaction function and the presence of first-order components are indicated for each model. The RMSE is computed from the three residual measures: raw $(\mathscr{R})$, Pearson $(\mathscr{R}^{\star k}_m)$, and inverse $(\mathscr{R}^{\dag k}_m)$.}
\end{table}
Table \ref{tab:RMSE} shows that the different types of residuals do not agree on a single best model. However, we can highlight our base model (Fiksel 1) as the best overall since it maintains low RMSE values across the other models. The model that assumes no interaction between cells of different types performs better regarding raw residuals; however, the Pearson and inverse residuals do not support this finding as much. This may be due to the variability in Figure \ref{fig:boxplotmodels}. The Strauss Hardcore model is highly competitive; the Pearson and inverse residuals favour this model despite having a slightly higher RMSE based on raw residuals than the Fiksel 1 model.

\section{Discussion}\label{sec:discussion}
In this article, we have proposed a multitype Fiksel interaction model for tumour immune microenvironments and applied it to understand inhomogeneity and interaction patterns of a sample of digitalised tissues through digital pathology techniques from 122 patients with lung cancer. 

Throughout this article, we have explored various tools connected through a statistical model that includes several components, a first-order component also called a trend that includes, in turn, the estimate of the expected value of the number of cells in each one of the tumour immune microenvironments through a non-parametric kernel and several design covariates. We have included all possible interactions between cells of the same type and cells of different types in a single component that describes the interaction; this term is a Fiksel-type pairwise interaction function, and it comes from the Gibbs and Markov pairwise interaction processes \citep{vanLieshout2000markov}. In summary, we have shown that inhomogeneous multitype Gibbs processes provide effective tools for analysing tumour immune microenvironments. It is the first time this multitype version has been used in practice since only the bivariate version was initially proposed \citep{fiksel1986pairwiseants}. 

Given that the images processed in digital pathology are relatively new, little has been studied from a statistical and probabilistic perspective on the distribution of immune cells within the tumour immune microenvironment \citep{wilson2021oportunitiesmultiplexdata}. Then, some open questions could give rise to new and exciting research fields. For example, are there asymmetric interactions between cell types? In other words, does a kind of cells appear or are located within the tumour immune microenvironment first, and then the other types are distributed conditionally to the first type? Hierarchical interaction models might account for this type of cell behaviour and assign, for example, our multitype Fiksel interaction function in a conditional way. A conditional hierarchical model would express the probability function such that
$$
f(X)=f_1(X_1)f_{2|1}(X_2|X_1)f_{3|1,2}(X_3|X_1,X_2)\cdots f_{M|1,2,\ldots,M-1}(X_M|X_1,\ldots,X_{M-1}),
$$
where $X=\cup_{m\in \mathscr{M}}{X_m}$ has $M$ point types, and $X_{m-1}$ takes precedence over $X_m$ for every $m\in \mathscr{M}$. Each probability density $f_{m|1,2,\ldots,m-1}$ is a pairwise interaction density that assembles all the information about the preceding terms, including the normalising constant, the trend and the interaction function \citep{hogmanderandsarkka1999hierarchicalmultitype, grabarnik2009hierachical, baddeley2015spatialR}. 

On the other hand, the alternative models (see Section \ref{sec:severalmodels}), particularly Fiksel 2 and 3, have shown that the assumption of homogeneity could be reasonable in this context. Unfortunately, a homogeneity test through quadrat counting \citep{baddeley2015spatialR} would be inadequate given the dependency between cells; therefore, we cannot formally rule this homogeneity out. Nevertheless, although the first-order factors are statistically significant (see Section \ref{sec:regularparameters}), the performance of these models remains similar to those that include first-order terms. Therefore, a homogeneous, more parsimonious model, such as the ones we have offered, may be adequate in this case.

It is important to highlight that our estimates are point estimations and could be improved. for example, through simulation using {\it Metropolis-Hastings algorithms} for Gibbs processes \citep[see, e.g., \ ][chap. 7]{moller2004}. These algorithms can provide confidence intervals for the associated parameters. As an interesting future research direction, the proposed model also offers a good opportunity to estimate parameters using other approximate Bayesian computational methods, such as the variational Bayesian method \citep{ren2011variational}. 

One of the problems we face is the amount of data that results when applying the procedures of generated linear models computationally. Although we have very well-optimised software nowadays, sometimes it is not enough. In our case, we have about seven million records that the regression algorithm must process and the other point process techniques described throughout the paper, such as kernel smoothing and K-function calculation, which fortunately were calculated only for each patient. The problem we have faced goes beyond any processing speed; the problem is the vast amount of memory required to do all the calculations, which requires serious computational resources. That is why an interesting line of research could include how to bring these computations in the context of this type of digital pathology images to the comfort of a conventional laptop. 

We conclude that multitype inhomogeneous Gibbs models are a convenient statistical option for tumour immune microenvironment analysis. In particular, the Fiksel interaction function is satisfactory for studying the interaction between cells of the tumour immune microenvironment. These models can easily include extra clinical information available per individual, although they are robust enough to provide good results even without this first-order information. These models also allow estimation and inference through the computational simplification offered by pseudolikelihood methods.

\bibliographystyle{agsm}
\bibliography{bibliography}
\end{document}